\documentclass[12pt,a4paper]{article}
\usepackage[utf8]{inputenc}
\usepackage[english]{babel}
\usepackage{amsmath}
\usepackage{amsfonts}
\usepackage{amssymb}
\usepackage{makeidx}
\usepackage{graphicx}
\usepackage{morefloats}
\usepackage[left=2cm,right=2cm,top=2cm,bottom=2cm]{geometry}
\author{Arindam Ghosh$^{1}$ and Sandip K. Chakrabarti$^{1,2}$}
\title{Signature of Two-Component Advective Flow in several Black Hole candidates obtained through time-of-arrival
analysis of RXTE/ASM Data}
\begin{document}
\maketitle
\begin{center}
{$^1$S. N. Bose National Centre for Basic Sciences, Salt Lake, Kolkata 700106, India.\\ 
$^2$Indian Centre for Space Physics, Chalantika 43, Garia Station Rd., Kolkata 700084, India.\\
{\it arindam.ghosh@bose.res.in; chakraba@bose.res.in}\\}
\end{center}
\abstract{We study several Galactic black hole candidates using long-time RXTE/ASM X-ray data to search for telltale signatures of differences in viscous timescales in the two components used in the Two-Component Advective Flow (TCAF) paradigm. In high-mass X-ray binaries (HMXBs) mainly winds are accreted. This nearly inviscid and dominant sub-Keplerian flow falls almost freely towards the black hole. A standard Keplerian disc can form out of this sub-Keplerian matter in presence of a significant viscosity and could be small in size.
However, in low-mass X-ray binaries (LMXBs), highly viscous and larger Keplerian accretion disc is expected to form inside the sub-Keplerian disc due to the Roche-lobe overflow. Due to two viscous timescales in these two components, it is expected 
to have a larger lag between the times-of-arrival of these components in LMXBs than that in HMXBs. Direct cross-correlation between the photon fluxes will not reveal this lag since they lack linear dependence; however, they are coupled through the viscous processes which bring in both matter. To quantify the aforesaid time lag, we introduce an index ($\Theta$), which is a proxy of the usual photon index ($\Gamma$). Thus, when $\Theta$, being dynamically responsive to both fluxes, is considered as a reference, the arrival time lag between the two fluxes in LMXBs is found to be much larger than that in HMXBs. Our result establishes the presence of two dynamical components in accretion and shows that the Keplerian disc size indeed is smaller in HMXBs as compared to that in LMXBs.}\\ 




\section{Introduction}
X-rays are emitted from accretion discs around black holes primarily through two mechanisms. An optically thick but geometrically thin, Keplerian disc dissipates gravitational energy through viscosity and radiates multi-colour blackbody (soft) photons (Shakura \& Sunyaev 1973; hereafter SS73). 
Another component is a hot electron cloud which emits a non-thermal power-law after it intercepts soft photons from the disc.
In a self-consistent theoretical paradigm, called a Two Component Advective Flow (TCAF) solution (Chakrabarti \& Titarchuk 1995; hereafter CT95), the latter component is the result of a relatively hotter, optically thin but geometrically thick, sub-Keplerian halo which sandwiches the Keplerian disc and dissipates its potential energy through repeated inverse Comptonization. In TCAF paradigm, the two components are due to differential viscosity in the vertical direction. Equatorial regions with viscosity parameter higher than the critical values will form a Keplerian standard disc (Chakrabarti 1990, 1996), while the rest would remain sub-Keplerian, falling rapidly towards the black hole. This dynamic halo slows down close to the horizon due to centrifugal barrier and forms a standing or oscillating shock wave. The post-shock region behaves as a Compton cloud which intercepts photons from the Keplerian disc located in the pre-shock region on the equatorial plane and 
inverse Comptonizes them following prescriptions of Sunyaev \& Titarchuk (1980). TCAF paradigm generally addresses all the observational aspects of a black hole candidate quite satisfactorily.

In the literature, several models were presented which use a `hot electron cloud'  or a `static corona' to produce the power-law component
of the spectrum (Sunyaev \& Titarchuk 1980, 1985; Zdziarski 1988; Liang \& Wandel 1991, Haardt \&  Maraschi 1991, 1993; Haardt et al. 1994; Meyer \& Meyer-Hofmeister 1994; Narayan \& Yi 1994; Liu et al. 1999, 2002; Meyer et al. 2000; Taam et al. 2008; Zhang et al. 2009; Poutanen et al. 2018). 
These models make varied assumptions about the disc and the corona and often are successful in explaining specific aspects of observations.
In this paper, we shall follow the TCAF paradigm, because it requires a minimal number of parameters while fitting
spectra of black hole accretion and also because it is derived out of rigorous theory of transonic accretion flows. Most importantly, a TCAF
configuration is found to be stable. Although the Keplerian component can be produced from the sub-Keplerian component due to a rise
in viscosity farther away, the two components behave independently closer to the black hole as the numerical simulations of 
Giri \& Chakrabarti (2013) have shown. At the outer edge, there could be a single component, as in a high mass X-ray binary (HXMB), 
where the flow is mainly from the winds of the companion or, a mixture of both the components as in a low mass X-ray binary (LMXB).
Interestingly, viscosity is explicitly never required. It is manifested in segregating two components of independent accretion rates.

In TCAF, there are four free parameters, viz. the two mass accretion rates, shock location, and shock strength. Soft flux mainly depends on the disc rate, but the hard flux depends on all three parameters. The variation in X-ray fluxes emitted from the disc and the halo, and in spectral index directly reflects the change 
in the mass accretion rates of the two components close to the black hole. Although  viscosity is not explicitly required to fit an instantaneous spectrum, for studies of time evolution, its effects must be considered, as after all, the disc matter is ushered in by viscous processes.
If the viscosity were constant for both the flows, the time variation at the outer edge would have been 
reflected at the inner edge of the disc after a constant viscous time delay. 
However, an SS73 disc requires a higher transport rate of angular momentum to maintain the Keplerian distribution,
and thus is expected to have a higher viscosity parameter, especially that the overall disc pressure is low. 
On the other hand, the advective component having low angular momentum 
requires very little viscosity for accretion, especially that its ion pressure is high. 
As a result, out of the two components, the advective halo component arrives earlier than the Keplerian viscous component. 
In an outburst source, the time lag between these two components directly depends on the pile-up radius of matter,
from which the matter components rush to the black hole in respective timescales when high viscosity triggers an outburst. Thus the lag could be different in different outbursts.
The hot electron cloud in CT95 is created after the radiatively inefficient advective flow having 
low angular momentum faces a centrifugal barrier close to the black hole, where a reduction of radial kinetic energy 
results in heating and puffing up of the flow. This CENtrifugal pressure supported BOundary Layer (CENBOL) 
represents the Compton cloud and converts intercepted 
soft photons into hard photons. The CENBOL is bounded outside by the shock. Rigorous analysis of real data reveals that the TCAF theory can explain even subtle aspects of temporal as well as spectral properties of black holes, including quasi-periodic oscillations (QPOs), outflow-rate, and spectral state correlation,  etc. (Chakrabarti \& Manickam 2000; Rao et al. 2000; Smith et al. 2001, 2002a-b, 2007; Wu et al. 2002; Debnath et al. 2010; Soria et al. 2011; Nandi et al. 2012; Cambier \& Smith 2013). Mass of the black hole is also calculated accurately from each observation. The Compton cloud being a natural consequence of the TCAF solution, no separate hot electron cloud is required. The sub-Keplerian flow, forms due to winds of the companion star, arrives nearly in free-fall timescale. An increase in halo rate will harden the spectral index without affecting the soft or hard photon flux significantly. Smith et al. (2001, 2002b) studied this aspect using both RXTE/ASM (3-12 keV) and RXTE/PCA (2.5-25 keV) data and found that an assumption of the existence of two components in accretion flows would allow one to interpret the results far better than a single component flow such as the advection-dominated flow (Narayan \& Yi 1994), or the persistent thin disc (Mineshige 1996). They found that there was a distinct time lag between a suitably defined power-law index and hard photon flux in LMXBs, such as GRS 1758-258 and 1E 1740.7-2942. However, HMXBs, such as Cyg X-1 and Cyg X-3, do not show any significant lag.\\   

The viscous timescale in an accretion flow, $t_{vis}$, may be written as (Frank, King, \& Raine 2002)
$$
t_{vis}=\frac{r^2}{\alpha c_s z},
$$
where, $\alpha$ is the viscosity parameter prescribed in SS73 ($\alpha <1$), $c_s$ is the speed of sound, $z$ is the height of the disc at a radial distance $r$ away from the black hole. If any of these four quantities varies, $t_{vis}$ will change. $t_{vis,disc}$ could remain constant for relatively smaller discs, provided the unknown viscosity parameter $\alpha$ or pressure does not fluctuate appreciably. This is likely to be the case for HMXBs. If the above equation is true, then a high value of $\alpha$ in bigger discs, such as in LMXBs, may result in small $t_{vis,disc}$. However, both $\alpha$ and viscous stress could vary spatially and temporally. On the contrary, $t_{vis,halo}$ of sub-Keplerian flow (halo), with its  negligibly small viscosity, is roughly comparable to the free-fall time $t_{ff}$. Therefore, a difference (true lag) in two timescales is expected. 
Since there is no possibility for $t_{vis,disc}$ to be less than $t_{vis,halo}$, we unambiguously define, 
$$
\tau = t_{vis,disc}-t_{vis,halo} \approx t_{vis,disc}-t_{ff}.
$$
Thus, $\tau>0$ will imply that the disc lags behind the halo. This is an important consequence of the TCAF theory, and is generic by nature. 
If $\tau=0$, it means that the disc evolves 
simultaneously with the halo. This may happen, even if temporarily, before an outburst is triggered in transient sources. 
In a recent work (Roy \& Chakrabarti 2017), 
it is shown by an extensive simulation that a rise in $\alpha (r,z,t)$ may trigger an outburst where 
the Keplerian disc rate goes up significantly. However, in different outbursts of the same source, the accumulation radius (or, pile-up radius) of matter 
as well as $\alpha$ may be different. Thus the time lag between the peaks of the Keplerian and the sub-Keplerian rate 
could be different from one outburst to another.\\
Inside the CENBOL, the sub-Keplerian flow and the Keplerian flow both become subsonic and indeed, due to turbulence arising out of back flow 
of matter after hitting the centrifugal barrier, the two timescales could become comparable, and the difference between them can be negligible. 
So, the time lag occurs in the pre-shock flows upstream before the creation of the Compton cloud.   
Even without any knowledge of the mass accretion rates, $\dot m_{d}$ and $\dot m_{h}$, a relative time difference ($\tau_{r}$) between the changes 
in soft and hard fluxes will give the true lag $\tau (\geqslant 0)$ between the two accretion rates. A sudden increase in $\dot m_{d}$ can enhance 
the soft flux (and instantaneously  the hard flux through inverse Comptonization of the resulting enhanced intercepted soft photons by CENBOL) in large discs.  So the time lag may be negligible.
Therefore, a lag ($\tau_{r}>0$) or no lag ($\tau_{r} \sim 0$ for a small disc) between two photon fluxes is likely to be observed at the onset of an outburst. Since during the outbursts,
ASM data is more accurate, we use the data during outbursts, whenever available, even though the issues we are raising are valid for any fluctuation in rates made at 
some outer radius. In a given outburst of an LMXB, if the disc is large, $\tau_{r}\gg0$ is expected. In smaller discs $\tau_{r}\geqslant0$ is expected.
In the former case, at the end of the peak outburst, the disc culminates in hard intermediate state before the hard state is achieved. In the latter case, the disc quite possibly gets accreted as soon as the peak outburst is over, and hard state may occur directly.\\  

In this paper, we explore the general lag characteristics of TCAF solution from the RXTE/ASM data. Our goal is mainly to show that 
both components move independently at two different timescales whence they are segregated. We show that specific accretion rate perturbations, which 
could have triggered simultaneously at some outer radius, take different amounts of time to reach the inner edge.  For instance, the increase 
in the halo component rate would harden the spectra without changing the hard or soft flux 
abruptly. However, only when the disc matter arrives close to the CENBOL, it cools CENBOL down rapidly 
and softens the spectra. To prove our point, we study several X-ray binaries falling under various categories,  
such as transient, persistent, and class-variable, using RXTE/ASM 3-channel data. We find that the conclusions regarding the time lags
drawn using a smaller data set earlier by Smith et al. (2001, 2002) with a different method, remain valid even for a larger data set. 
We show that the black holes in LMXBs indeed have, in general, larger accretion discs than those in HMXBs.
Different outbursts may have different disc size for LXMBs, mainly because the outer edge of the disc depends on the leftover viscosity after the previous event. Thus an LMXB may have smaller discs in some outbursts; but in an HMXB, there is no Keplerian disc from the outer edge since the injected matter is sub-Keplerian. So the lags are always short.
Indeed we show in transient sources that the accumulation (pile-up) radius, from where the matter rushes 
to the black hole due to thermal-viscous instability, may be different in different outbursts in LMXBs, 
causing the time lag also variable for the same source. In the next Section, we present the 
procedure of Data Analysis which we employed in this paper. In \S3, we present the main
results of our analysis. Finally, in \S4, we summarize our results and draw conclusions.
  
\section{Data Analysis}
We use All Sky Monitor (ASM) data (1.5-12 keV) of Rossi X-ray Timing Explorer (RXTE) satellite. ASCII versions of public/archival RXTE/ASM daily-averaged lightcurve data from MJD 50500 (February, 1997) onwards, for the X-ray binary sources have been used. We analyze five transient sources, viz., GX 339-4, 4U 1543-47, XTE J1550-564, XTE J1650-500, \&  GRO J1655-40. Besides, two well-known sources, the persistent (HMXB) Cyg X-1 and the class-variable (LMXB) GRS 1915+105, are also analyzed with the data acquired over 13 years.\\
The RXTE/ASM has a collecting area of $90$ cm$^2$ and operates over $1.5-12~keV$ energy range. It consists of three channels, with three energy bands, viz. $A=(1.5-3~ keV)$, $B=(3-5~ keV)$, \& $C=(5-12~ keV)$ respectively. If $a$, $b$, \& $c$ are the number of photons in A, B, \& C bands respectively, then the A-band represents low-energy photons $a$ (or soft flux) with absorbtion, B \& C bands together represent the total Comptonized, high-energy ($3-12~ keV$) photon counts (or, hard flux) of $(b+c)$. For Cyg X-1 and GRS 1915+105, two hardness ratios, defined here as  $\mathrm{HR\mbox{-}1}=(b/a)$ and $\mathrm{HR\mbox{-}2}=(c/b)$ are examined for an initial guess on their spectral states and also for a comparison with previous reports. 

As the Comptonization efficiency (CE), defined by the ratio
$$CE=\frac{(b+c)}{a}\equiv \mathrm{HR\mbox{-}1}{(1+\mathrm{HR\mbox{-}2})},$$ 
includes information about both hardness ratios, 
it is computed for all the sources. We define $\Theta$ and compute it for each observation:
$$
\Theta = \tan^{-1} [\frac{(c-b)}{4.5}],
\eqno{(1)}
$$
where, the scale factor of $4.5$ arises from the difference in mean energies of C \& B bands. 
Here, $\Theta$ is really the slope at the harder photons of the spectrum using the {\it linear values} and 
not the logarithmic values of the counts or energy. That way $\Theta$ would be more sensitive to instantaneous variations in fluxes. 
With usual definition of power-law photon index $\Gamma$, namely, $\phi(E) \sim E^{-\Gamma}$,  
a slope of $\Gamma \leq 2$ would indicate a harder state while $\Gamma > 2$ would indicate a softer state. 
This is the slope in the $log (\phi(E))$ vs. $ log E$ plot. Thus, 
$$
dlog(\phi(E))/dlog E=-\Gamma . 
\eqno{(2)}
$$ 
However, we use $tan \Theta$=$d\phi(E)/dE=-\Gamma \phi(E)/E$ to define the spectral state, since 
ASM count could be negative and log scale cannot be used. We have,
$$
\phi \sim (-\frac{tan \Theta}{\Gamma})^{\beta} ,
\eqno{(3)}
$$
where, $\beta = \Gamma/({\Gamma+1})$. By definition, 
$-1.57<\Theta<1.57$, however, in reality $tan\Theta$ should be negative only, since $c<b$ is expected 
for equal energy binned data, which is not the case in the ASM data. Note that in softer states, 
$\phi(E)$ is large and $\Gamma\rightarrow 4$; $\beta \rightarrow 0.8$,
hence, $\Theta \rightarrow -1.57$. In our analysis, $\Theta_{lowest}$ is found to be around $-1.50$ 
(see Figs. 6h and 7). In the hardest states, it is expected that $\phi(E)$ is low and $\beta \rightarrow 0$,
$\phi$ tends to be a constant with $\Theta \rightarrow 0$. 
In ASM data, the width of the C band is the largest, so $c>b$ is possible 
and Eq. 1 often gives $\Theta>0$. One might expect that a direct cross-correlation between the 
hard flux and the soft flux $t_{(Hard,Soft)}\geqslant\tau$ would give the desired lag. However, 
this is not true since the hard flux is not dependent on the soft flux alone. Our main goal is to obtain the lag between the two accretion rates, but they do not independently
emit the two components of radiation either. Particularly, the hard flux is a function of the size and density of the CENBOL
as well as the soft flux. Second, hard flux starts rising as soon as the soft flux rises when viscous matter starts to move into the disc from piling radius. Thus one has to use $\Theta$ which is the dynamical slope and this gives the rate of enhancements of one component with respect to the other.  
Since, $\Theta$ varies continuously with 
the hard flux and the soft flux separately, we can compare one correlation with respect to the other for obtaining the viscous time lag between the rates.
If $t_{(\Theta,Soft)}$ and $t_{(\Theta,Hard)}$ are the outputs of cross-correlations of $\Theta$ with $Soft$ flux and $Hard$ flux respectively, then the relative time lag between the two fluxes is given by
$$
\tau_{r}=t_{(\Theta,Soft)} - t_{(\Theta,Hard)}\rightarrow\tau.
$$          
$\tau_{r}>0$ implies that the disc flux lags behind the halo radiation. If $\tau_{r}=0$, both evolve simultaneously.\\
In order to observe both the long-time and the short-time behaviours of the aforesaid sources, RXTE/ASM daily data for more than a decade is used. RXTE/ASM did not record X-ray data continuously from any particular source. Hence there are gaps in the data, especially, due to the annual solar constraints. The unevenness of data interval is filled up by using a {\sc fortran} code for 
interpolating data at 1d time interval. Using the {\sc perl} script {\sc ascii2flc}, the {\sc fits} files of the daily light curves are created. These are used to produce the cross-correlation between $\Theta$ and two fluxes using the {\sc crosscor} task of HEASOFT/XRONOS package.\\

\section{Results of our Analysis}

In order to show differences in time lag properties in various types of compact binaries, we present the results of the fluxes and correlations for various sources. $\Theta$ is the slope at the harder energies of the ASM spectrum (Eq. 1). With usual definition in logarithmic scale, a slope (in log-log scale) of $\le 2$ would indicate a harder state while $>2$ would indicate a softer state. We  use a convenient definition while handling ASM data and our $\Theta$ varies from $-1.57$ (softest) to $0$ (hardest). $\Theta>0$ represents points when $c>b$, an artefact of ASM data.

Transient sources which exhibit outbursts are very interesting in that they are irregular, and when the spectral data is fitted with TCAF
solution, the accretion rate in the disc is found to peak after a few days of peaking of the halo rate 
(Debnath et al. 2014; Mondal et al. 2014; Jana et al. 2016; Molla et al. 2016), the time gap being different for different
outbursts. Furthermore, shapes and energy release of each outburst of a single source could be totally different indicating 
that the accumulation radius (i.e., where the Keplerian disc matter was piling up before being 
released due to high viscosity) itself is different for each outburst. However, when ASM data is available and detailed spectral fit 
is not possible on a daily basis, we use the crude spectral slope $\Theta$ to establish the correlation.
Lightcurves for 5000 days of five transient sources, namely, GX 339-4, 4U 1543-47, XTE J1550-564, XTE J1650-500, \& GRO J1655-40 are drawn in Fig. 1 after taking daily average of the total or sum-band (1.5-12 keV) ASM counts. Figure 2 shows the variation of $\Theta$ with time; weekly running averaged data are used for clarity. Strong dips indicate the outbursts. In Figs. 1\&2, we mark the outburst events by OB followed by a number when multiple outbursts are present. We now concentrate in detail on these outbursts. In Fig. 3, we plot $\Theta$ (continuous/online-black curves) and CE (dotted/online-red curves) during the outbursts shown in Figs. 1\&2. Both quantities fall significantly just after the commencement of an outburst and rise again at the end of the outbursts. This is because at the onset of the rising phase and towards the end of a declining phase of any outburst, the state is hard/hard-intermediate and CE (Pal, Chakrabarti, \& Nandi 2013; Pal \& Chakrabarti 2015) is highest as the CENBOL is very large. Both the effective counts start to grow at different rates. That does not imply that they are correlated.
Note that at the beginning and at the end of each outburst, CE is the highest, though weekly averaged plots show only a few points and the  peaks are often smoothed out. $\Theta$ is also higher there since the spectral index is flatter for harder states as the existing spectrum gets hardened when the advective sub-Keplerian matter rushes in. Similarly, as the Keplerian matter rushes in to increase the soft flux, the power-law part is softened. As a consequence, the CENBOL cools down and starts to collapse; both $\Theta$ and CE go down, giving rise to softer states. Thus both CE and $\Theta$ are changing continuously. We thus concentrate on these two quantities. Hence it is the slope of the two components (represented by $\Theta$) which must be correlated with both the fluxes independently.

Five outbursts in GX 339-4 (of Figs. 1a, 2a, \& 3(I)-3(V)) and cross-correlation profiles are shown in Fig. 4. Figure 5 shows the results of four outbursts in four sources, viz. 4U 1543-47, XTE J1550-564, XTE J1650-500, and GRO J1655-40 (Ref. Figs. 1b-1e, 2b-2e, 3(VI)-3(VII) \& 3(X)). In all the cases, hard and soft fluxes are respectively drawn by dotted/online-blue and continuous/online-red lightcurves; dotted/online-blue and continuous/online-red curves in the correlation profiles represent correlations of $\Theta$ with hard-flux and soft-flux respectively. $\Theta$ and flux-flux correlation curves are drawn in dot-dashed/online-black. The correlation profile with zero-day lag in OB1 (declining phase) of GRO J1655-40 is not shown. In view of Figs. 3(II), 3(V), 3(VIII), 4(b1), 4(e1), and 5(c1), it is seen clearly that when the time lag is high, as obtained from $\Theta$-flux cross-correlations [Figs. 4(b2), 4(e2), and 5(c2)], it is reflected from $\Theta$-behaviour alone. In particular, OB2 of GX 339-4 shows prominently a lag of 22d [Figs. 3(II) or 4(b1)], which is consistent with the time lag of 19d computed from $\Theta$-flux cross-correlations [Fig. 4(b2)]. In Fig. 3(II), $\Theta$ rises from its high (hard state) value (about $0$) at the commencement of OB2 to a higher (harder state) value (about $0.6$); the time elapsed in such a spectral hardening is the desired time lag.\\ 
At this juncture, in view of the plots in Figs. 4(a1)-(a5), 5(b1), 5(c1), \& 5(d1), we can classify the outbursts (OB) 
in two categories: OB-I and OB-II. Figures 4(a1), 4(c1), 5(b1), and 5(d1) conform to OB-I; while OB-II pertains to Figs. 4(b1), 4(d1), 4(e1), and 5(c1). In OB-I, both soft and hard fluxes increase in a similar manner throughout the duration of the outburst ($t_{OB}=t_{ris}+t_{dec}$). The duration of their rising phase ($t_{ris}$) is comparable to that of the declining phase ($t_{dec}$). For $t_{OB}>50d$ (say), $1< (t_{dec}/t_{ris})\leqslant 2$. From Figs. 3(I), 3(III), 3(VI), 3(VII), and 3(X) it appears that CE (and/or $\Theta$) fluctuates more about its mean highest value at the commencement of the outburst before dropping quickly to its lowest value. In OB-II, hard flux increases first and faster than the soft flux in the rising phase. The duration of their declining phase ($t_{dec}$) is longer than that of the rising phase ($t_{ris}$). For $t_{OB}>50d$ (say), $2< (t_{dec}/t_{ris})\leqslant 6$. It is readily observed from Figs. 3(II), 3(IV), 3(V), and 3(VIII) that at the onset of the outburst CE (and/or $\Theta$) becomes roughly steady (plateau region) about its mean highest value (also see Fig. 7 later) almost for a month, before falling off relatively slowly to its minimum. Outbursts of short duration ($t_{OB}<50d$, say), as that in 4U 1543-47 (Fig. 5(a1)), 
may belong to OB-I. In OB-I, $\tau_{r}\rightarrow 0d$ (Figs. 4(a2), 4(c2), 5(a2), 5(b2)); while in OB-II,
$\tau_{r}\gg 0d$ (Figs. 4(b2), 4(d2), 4(e2), 5(c2), 5(d2)).\\ 
We now plot in Fig. 6 the mean photon energy flux (mean intensity) in all ten outbursts discussed above as a 
function of $\Theta$. These are arranged anticlockwise in the descending order of $\tau_{r}$-values (written in each box) from the top left Fig. 6a to the top right Fig. 6j. We see that $-1.50 \leqslant \Theta \leqslant 1.25$ 
for all the cases, though $\Theta$ does not normally go above $\sim 0.5$. Wild fluctuations along $\Theta$-axis are due to poor counts in ASM data. There is a specific shape in the excursion path of the {\it hysteresis loop} (analogous to {\it q-track pattern} observed in an HR-intensity diagram). Most interestingly, the area within the loop appears to be roughly proportional to $\tau_{r}$. Outbursts with negligible $\tau_{r}$ also has a negligible loop area. This is why we preferred to analyze all outbursts as a whole instead of analyzing their rising and declining phases separately. In Roy \& Chakrabarti (2017) it was concluded that the hysteresis loop in the outbursts (i.e., inability to retrace the rising phase track by the declining phase) is due to inability of the Keplerian disc to dissipate itself quickly in the absence of viscosity after the soft state is achieved. A larger Keplerian disc would take a longer time to disappear. Thus the retracing track deviates more for a larger Keplerian disc. So a larger $\tau_{r}$ also means a larger Keplerian
disc and/or lower viscosity. Hence, $\tau_{r}$, together with the loop area, indicates the size of the 
Keplerian disc. Clearly, the size of the disc is different in different outbursts. The variations of $\Theta$ and CE with
$\tau_{r}$ obtained from all outbursts are depicted in Fig. 7. Mean CE and $\Theta_{mean}$ vary in a similar manner. It is seen that the maximum value of CE indefinitely fluctuates as $\tau_{r}\rightarrow0d$, which is obtained in OB-I type outbursts in LMXBs with small disc; although it becomes  approximately stable and constant for $\tau_{r}\gg 0d$, 
which is attributed to OB-II type outbursts in LMXBs with a big disc.\\

Figures 8 and 9 respectively show the long time behaviour of Cyg X-1 and GRS 1915+105 through
six quantities, viz. (a) hard flux, (b) soft flux, (c) HR-1, (d) HR-2, (e) CE, and (f) $\Theta$. Weekly running mean data are used. Note the considerable variabilities in all the time-series. Due to background subtraction effects, $\Theta$ is mostly positive. In regions where soft and hard fluxes are the lowest, CE is the highest and $\Theta$ is also the highest. This implies harder spectra. This is true in both objects. What is not obvious from Figs. 8 \& 9 is whether there is a definite time lag between the photon fluxes and $\Theta$. Their correlation profiles (using the seconds-data by interpolation at $0.1d$=$2.4~hour$ interval) are shown in Figs. 10a \& 10b. Colour stamps or line types mean the same as before. Reig, Papadakis, \& Kylafis (2002) obtained similar flux-flux cross-correlation profiles (without any lag) separately using the three bands of long-term RXTE/ASM data. We note that the responses of these two 
objects are instantaneous. Hard and soft fluxes have no relative lags. The correlations between 
hard and soft fluxes show no measurable lags also. 
Indeed, $\Theta$ and CE change simultaneously 
with both photon fluxes (Figs. 8e-8f, 9e-9f). 
So, there should be really a negligible lag in both Cyg X-1 and GRS 1915+105. This implies that the size of the Keplerian disc cannot be large.\\ 
Though Cyg X-1 is considered to be a persistent source, it very often flares up and reaches the soft state. 
It may be instructive to see if its zero-lag property of these flares is similar to the zero-lag outbursts of transient sources. 
We have already noted that the behaviour of $\Theta$ (Fig. 8f) and correlation profile (Fig. 10a) 
in Cyg X-1 is exactly similar as those of the outburst cases (Figs. 4 and 5), particularly to OB-I type (e.g., 4U 1543-47, and OB1 \& OB3 in GX 339-4).
This motivates us to look into six soft states (marked as i-vi in Fig. 8f). Interpolated data of every $2.4~hour$ are used so that $\tau_{r}>5~hour$ 
could be detected. However, not surprisingly, we find that their correlation profiles 
with $\tau_{r}=0$d, remain identical to that in Fig. 10a. Figure 11 shows the mean photon energy flux vs. $\Theta$ plots (hysteresis loops). Here too we see a similar excursion of data points. Being a wind-fed system, Cyg X-1 could exhibit {\it minuscule outburst}s or {\it flare}s. 
The size of the discs is different in different outbursts, but for persistent source Cyg X-1 such an ambiguity is absent. 
It seems that whenever its small disc is formed, 
it is drained out quickly and the hard state is returned.\\
All our results are summarized in Table 1 for comparison.    As expected, all the flux-flux correlations are seen to have $\sim 0d$ time lag.
 
\section{Summary}
Our detailed analysis of several stellar mass black hole candidates, 
which could be persistent, transient, or class-variable type, 
show very important lag properties. Since in any outburst, hard flux starts to change simultaneously with the soft flux, a direct correlation 
between the hard flux and the soft flux would not give us the lag. However,
because of two different time scales of rise of the halo and disc rates cause these photon fluxes to be coupled in a non-linear way. Thus, we attempted to find the lag using a third variable, namely, the spectral slope $\Theta$ (from a linear-linear plot rather than a log-log plot), with respect to which the fluxes were correlated. We see very clearly 
in all these sources that the soft fluxes either lag hard fluxes or these 
two change roughly simultaneously (no lag).  This could be easily interpreted when there are two components in the flow (CT95), as the advective 
component, which mainly controls the hard flux, always accretes
quicker than the Keplerian component which primarily controls 
the soft X-ray flux. The Keplerian disc lags because of its higher angular momentum which is transported in a longer viscous 
time scale. The advective component is of lower angular 
momentum and does not require, in general, any viscosity. Our work 
reveals that the two components in an accretion flow are separable 
with or without a lag in all sources considered here. In HMXBs, 
as well as in class-variable GRS 1915+105, both flows evolve almost 
simultaneously. Transient sources with either sporadic outbursts 
(for example, 4U 1543-47) or with regular outbursts 
(for example, GX 339-4) reveal that no lag is practically required in 
order to trigger an outburst in LMXBs 
with a small disc. However, a significant lag in  LMXBs with big disc prevails in most of the cases. 
Most interestingly, we find that the behaviour of CE prior to an outburst can give an indication of
whether the  outburst is going to last longer or would be completed quickly. We have shown that
rapid fluctuations in CE prior to an outburst of OB-I type with short duration connote lower counts and small disc
size. A smoother CE in OB-II, where the disc is larger, clearly indicates higher counts in both high and low energy bands and the
following outburst is going to be of longer duration. This aspect is being explored and will be reported elsewhere. Although all our attempts to detect a lag/lead in Cyg X-1 with both long-time and short-time data have gone in vain, 
our results  points to the possibility that Cyg X-1 could be a {\it missing link} 
between the transient and other sources.
Reig, Papadakis, \& Kylafis (2002) have reported aperiodic class-variability 
as well as zero-lag between the photon fluxes in both Cyg X-1 and GRS 1915+105. Cyg X-1 is either persistently transient 
or transiently class-variable
but it marginally misses something what we recognize as {\it outburst}s, 
and smaller {\it flare}s are observed instead. This explains that the viscosity somehow 
remains close to the threshold and makes and breaks the Kepleian disc in short time scales. 
In the literature, already Smith et al. (2001, 2002) mentioned that discs in HMXBs are smaller in size 
while the larger lags in LMXBs indicate a larger Keplerian disc. Their conclusions were drawn based on a few years of 
RXTE/ASM data and a different approach. Our present results with a data from the same instrument but over a span of about 13 years, and using a completely different method, also converge to the similar conclusions. However, we concentrated on outbursts also
and showed that the lag of the same source could be different depending on the accumulation radius. The total 
area in our (mean intensity, $\Theta$)-plane appears to be directly related to the lag. We continued our study 
of the variable source as well and came to the similar conclusion that GRS 1915+105 also has a smaller disc. In Ghosh \& 
Chakrabarti (2016), we inferred a similar result based on the effectiveness of the smearing of tidal perturbation
at the outer edge disc by the viscosity. We showed that the Fourier spectra (power density \& periodogram) of both RXTE/ASM (1.5-12 keV) and Swift/BAT (15-50 keV) long-term lightcurves gave very sharp quasi-orbital periods (QOPs) for HMXBs (e.g., Cyg X-1) and broad QOPs for LMXBs (e.g., XTE J1650-500, GRS 1758-258, \& 1E 1740.7-2942). Since these independent methods
point to the similar conclusions about the disc size, we believe that we are closer to understanding the contributions of the two components in different X-ray binary systems.
     
\section*{Acknowledgement}
The authors are indebted to NASA Archives for RXTE/ASM public data and facilities.

\begin{table}[h!]
\begin{center}
\caption{Time Lag Characteristics in Galactic Black Holes}
\begin{tabular}{c|c|c|c|c|}
\hline\noalign{\smallskip}
X-Ray Binary & Nature & Data & Lag $\tau_{r}$ (d) & Lag $\tau_{r}$ (d)\\
Black Hole&(LMXB/HMXB)& Type & ($\Theta$-Flux) & (Flux-Flux)\\
\hline
&&&&\\
4U 1543-47 & Transient & OB & $0\pm 1.4$ & $0 \pm 1$  \\
&&&&\\
\hline
&&&&\\
XTE J1550-564 & Transient & OB & $1\pm 1.4$ &  $0\pm 1$ \\
&&&&\\
\hline
&&&&\\
XTE J1650-500 &   Transient & OB & $10\pm 2.8$ &  $6\pm 2$ \\
&&&&\\
\hline
& & OB1 & $0\pm 1.4$ & $ 0\pm 1$ \\
GRO J1655-40 &Transient &&&\\
& & OB2 & $3\pm 1.4$ &  $0\pm 1$\\
\hline
 &  & OB1 & $0\pm 1.4$ & $ 0\pm 1$ \\
& & OB2 & $19\pm 1.4$ & $0\pm 1$\\
GX 339-4& Transient& OB3 & $1$  $\pm 1.4$ & $0\pm 1$ \\
& & OB4 & $7\pm 1.4$ & $3\pm 1$ \\
& & OB5 & $19\pm 4.1$ & $0\pm 1$\\
\hline
&&&&\\
Cyg X-1 & Persistent & Long Time & $0\pm 0.1$  &  $0\pm0.1$ \\
&&&&\\
\hline
&&&&\\
GRS 1915+105 & Class-Variable & Long Time & $0\pm 0.1$ & $0\pm 0.1$ \\
&&&&\\
\hline
\end{tabular}
\end{center}
\end{table}
\newpage 

\begin{figure}
\begin{center}
\includegraphics[height=18 cm, angle=0]{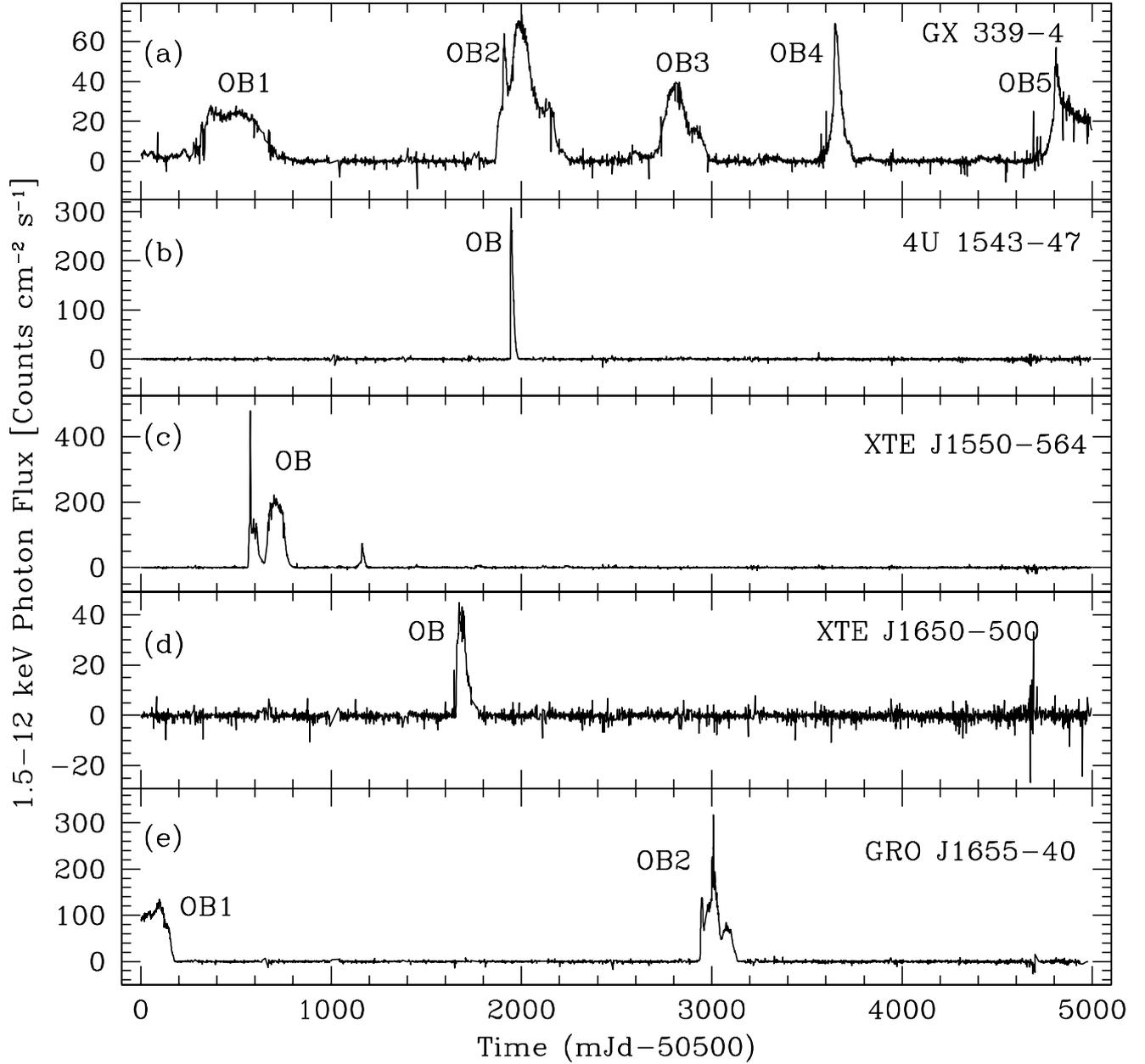}
\caption{(1.5-12 keV) lightcurves of five transient sources with long-time RXTE/ASM daily average data. The outbursts of our interest are abbreviated as OB, followed by serial numbers wherever multiple outbursts are present.}
\end{center}
\end{figure}
   
\begin{figure}
\begin{center}
\includegraphics[height=18 cm, angle=0]{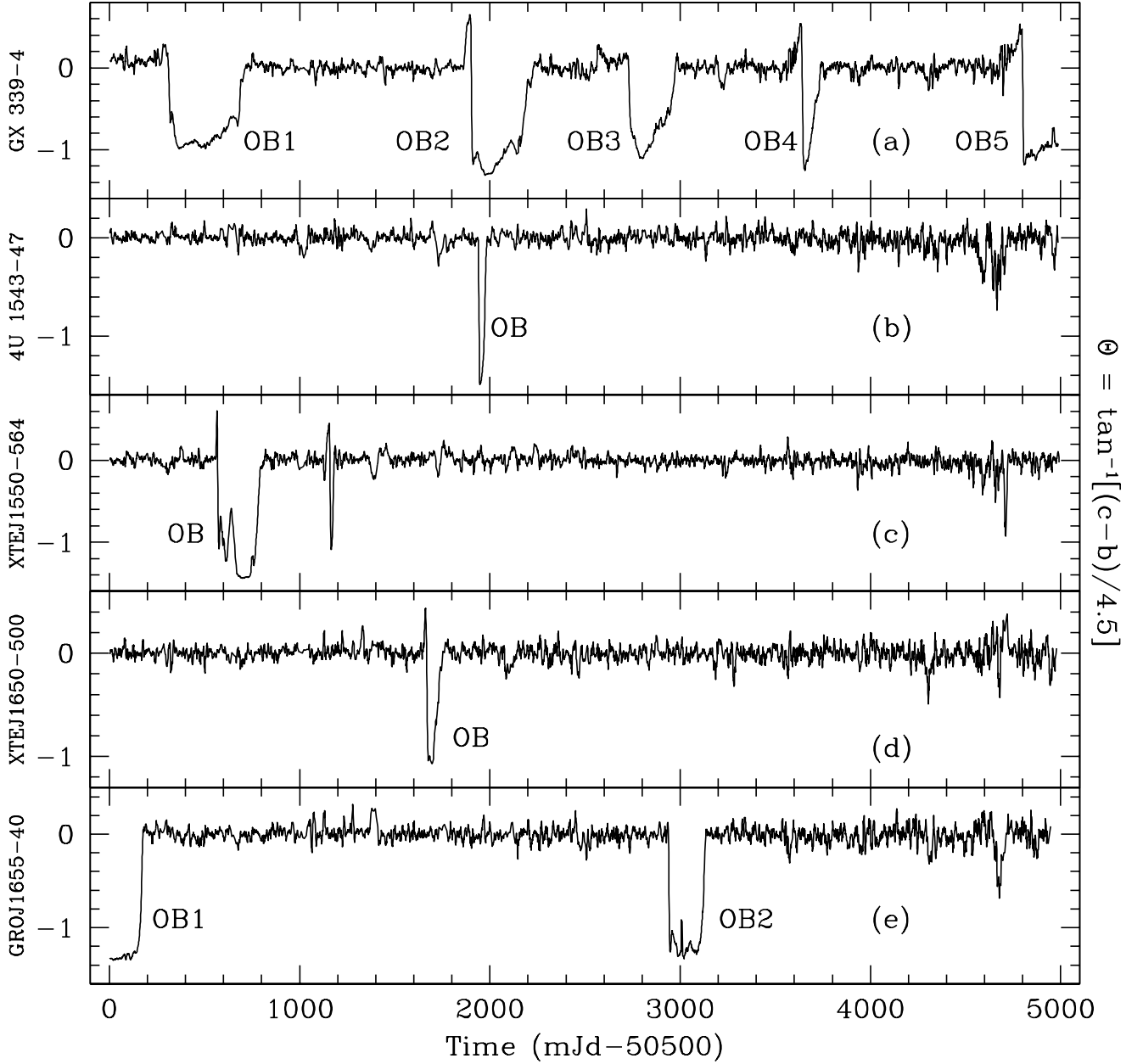}
\caption{Weekly running-mean $\Theta$ in the five transient sources is shown corresponding to Fig. 1. Strong dips are showing the outbursts, marked as OB and followed by serial numbers as before. As expected, $\Theta$ drops drastically at the commencement of an outburst and rises again at its end.}
\end{center}
\end{figure}
\begin{figure}
\begin{center}
\includegraphics[height=18 cm, angle=0]{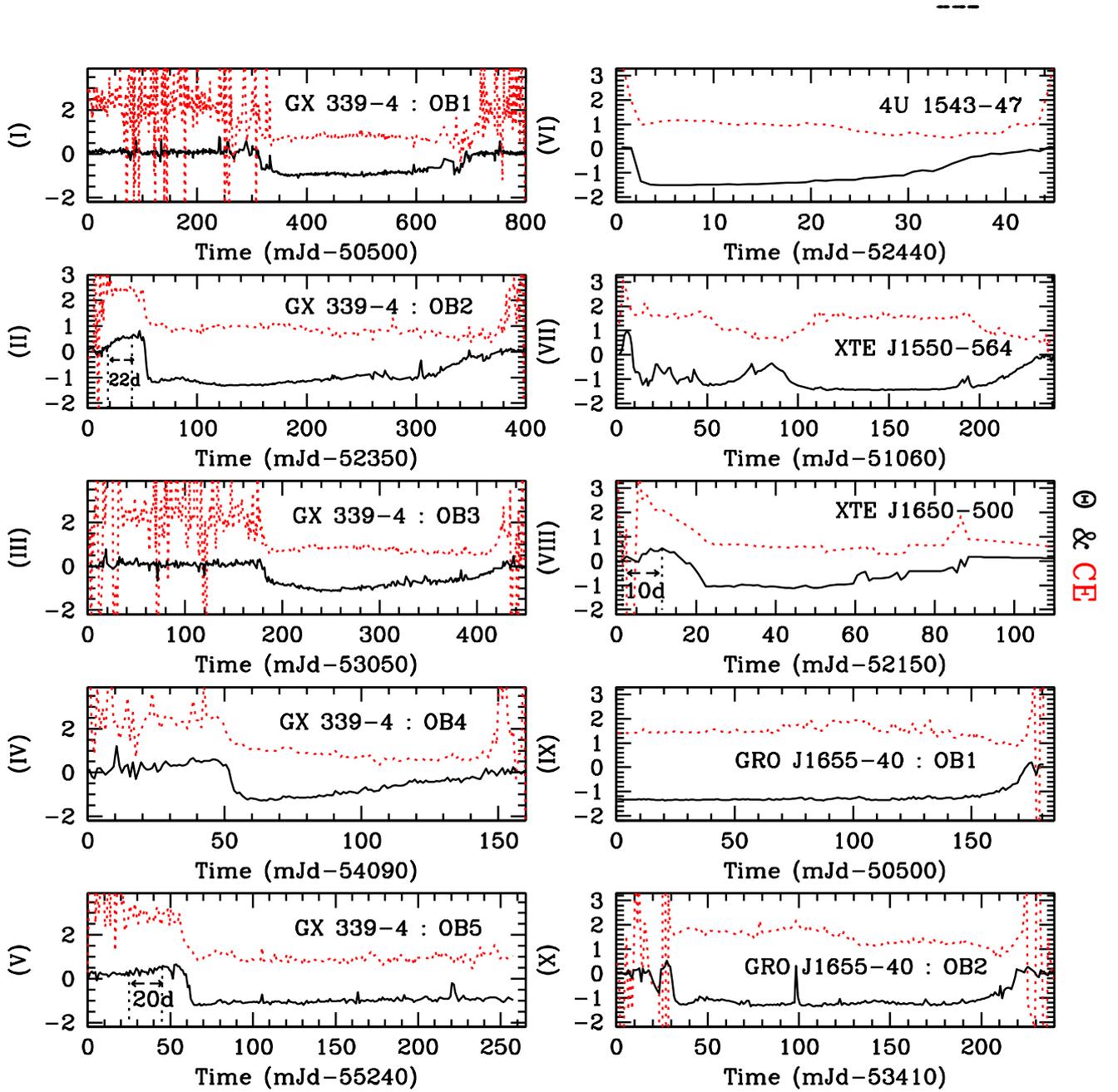}
\caption{Both $\Theta$ (continuous/online-black) and CE (dotted/online-red) are drawn with daily mean data during ten outbursts of Figs. 1 \& 2. 
Without any loss of generality, higher values of CE are ignored. 
This shows the physical significance of $\Theta$. Both $\Theta$ and CE fall 
significantly during outbursts, and rise again at the end of it. However, the approximate time elapsed during a spectral transition from hard to harder state, even after the commencement of the corresponding outburst and prior to the achievement of the soft state, is indicated in II, V, and VIII.}
\end{center}
\end{figure}
\begin{figure}
\begin{center}
\includegraphics[height=18 cm, angle=0]{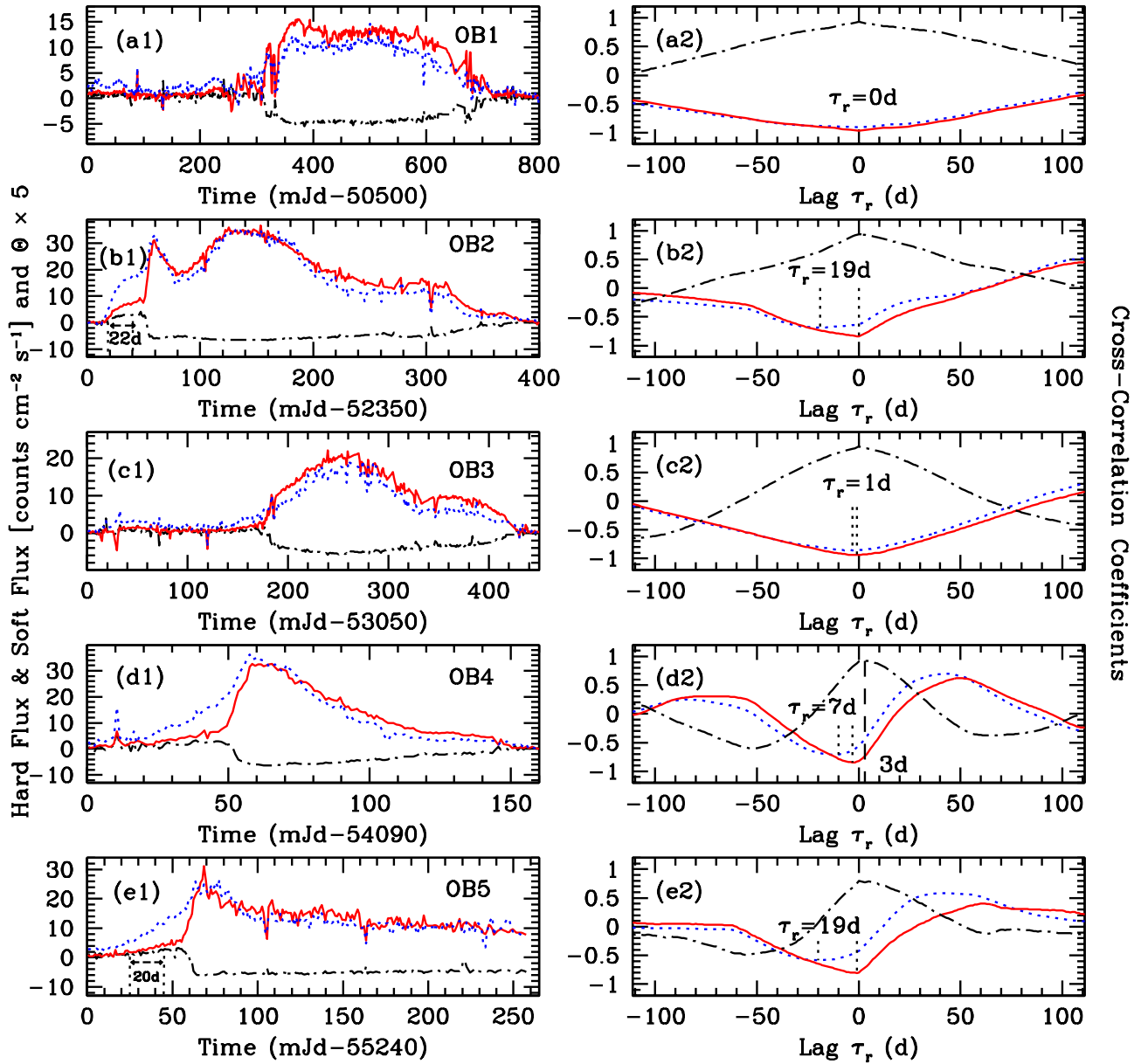}
\caption{Five outbursts of GX 339-4 [Ref. Figs. 1a, 2a, \& 3(I)-(V))]: Two fluxes and scaled $\Theta$ are shown on the left. Cross-correlation profiles are plotted on the right. Continuous/online-red and dotted/online-blue curves pertain to soft flux and hard flux respectively, while dot-dashed/online-black curves refer to $\Theta$ and flux-flux correlation. The relative lag $\tau_{r}$ values, obtained from $\Theta$-flux cross-correlations, are marked in each box. Flux-flux correlation results in $\tau_{r}=0d$ in (a2), (b2), (c2), and (e2); while  the lag time of $3d$ is marked in (d2). Time stamps labeled in (b1) \& (e1) are the approximate times of spectral state transition from hard to harder states. As expected, these are consistent with the time lags marked in (b2) \& (e2) respectively.}
\end{center}
\end{figure}
\begin{figure}
\begin{center}
\includegraphics[height=18 cm, angle=0]{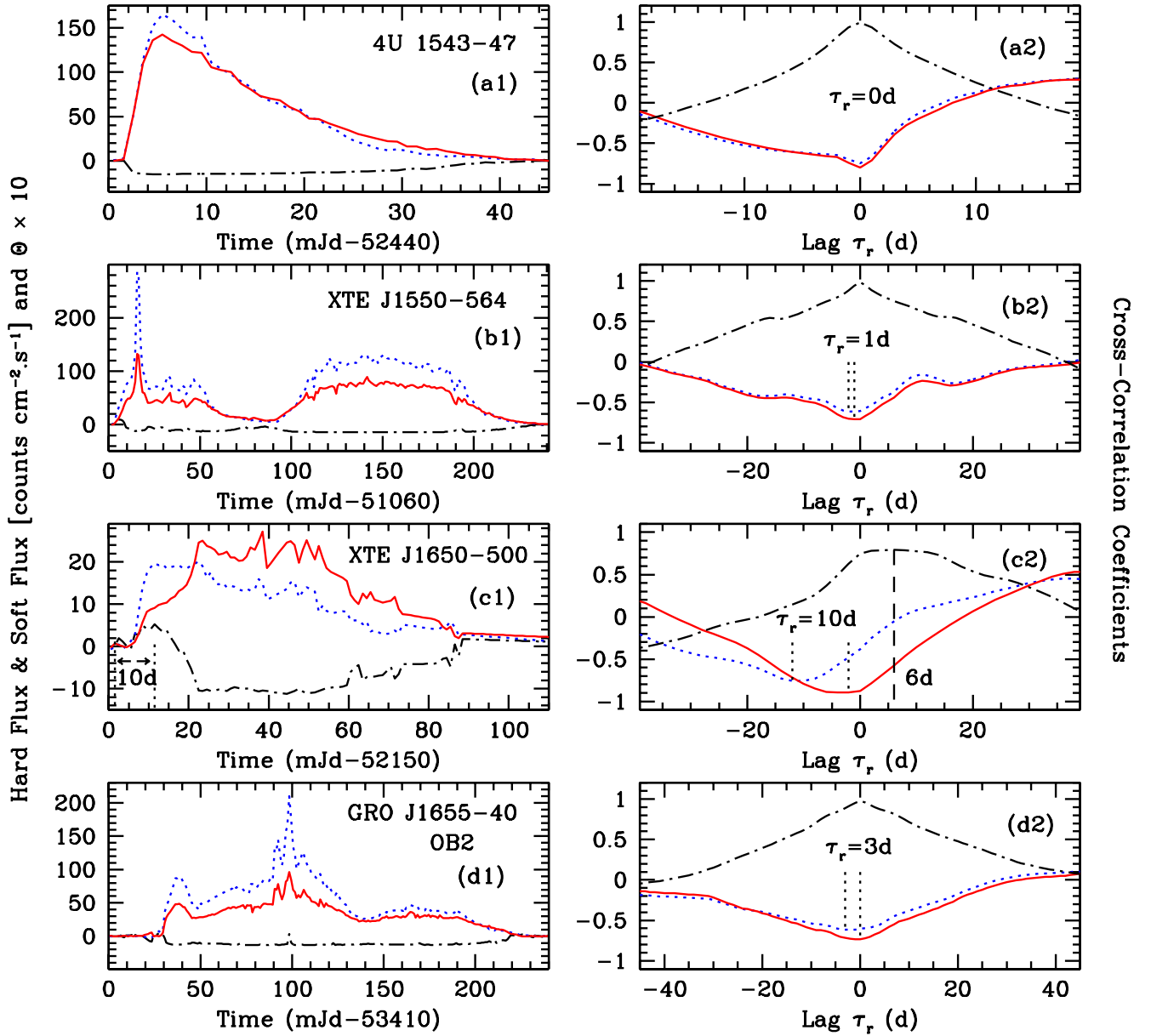}
\caption{Daily lightcurves and scaled $\Theta$ of the four sporadic outbursts in four transient sources 4U 1543-47, XTE J1550-564, XTE J1650-500 \& GRO J1655-40 are shown on the left [Ref. Figs. 1b-1e, 2b-2e, \& 3(VI)-(VIII) and 3(X)].  Corresponding correlation profiles are shown on the right. Continuous/online-red and dotted/online-blue curves pertain to soft flux and hard flux respectively, while dot-dashed/online-black curves refer to $\Theta$ and flux-flux correlation. The values of $\tau_{r}$, obtained from $\Theta$-flux cross-correlations, are marked in each box. Flux-flux correlation results in $\tau_{r}=0d$ in (a2), (b2), and (d2); but the lag time of $6d$ is indicated in (c2). The time stamp labeled in (c1) is the approximate time of spectral state transition from hard to harder state. This is consistent with the outcome in (c2).}
\end{center}
\end{figure}
\begin{figure}
\begin{center}
\includegraphics[height=18 cm, angle=0]{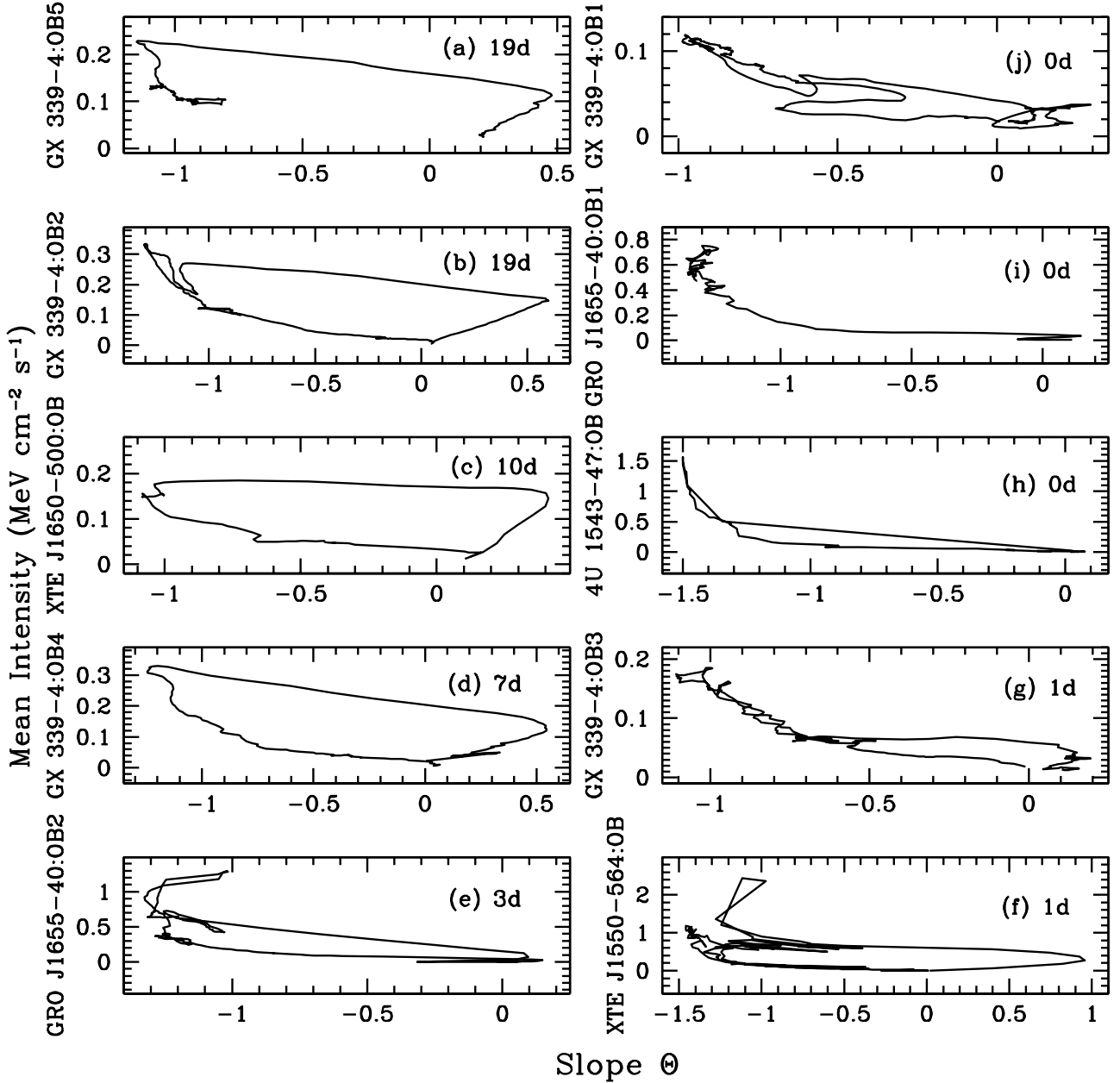}
\caption{Mean intensity in all the outbursts of Figs. 4 \& 5 is plotted with $\Theta$. This physically significant picture can  replace {\it q-track pattern} of any relevant HR-intensity diagram. $\tau_{r}$ values (of Figs. 4 \& 5) are marked in each box. (a)-(g) conform to $\tau_{r}>0d$, in descending order from the top left. The typical loop, if seen anticlockwise from (a),  gradually becomes thinner as $\tau_{r}$ disappears. As $\tau_{r}\rightarrow 0d$, the {\it hysterisis} loops become narrowed further in (g) to (j). Wild fluctuations in some cases are due to ASM data. In general, the loops on the left and the right are respectively pertaining to OB-II and OB-I types of outbursts. The incomplete loops in (a) and (i) are due to the fact that the outburst in (i) is a declining phase, whereas the outburst in (a) partially lacks this phase due to the choice of our data set (see Figs. 1a \& 1e).}
\end{center}
\end{figure}
\begin{figure}
\begin{center}
\includegraphics[height=15 cm, angle=0]{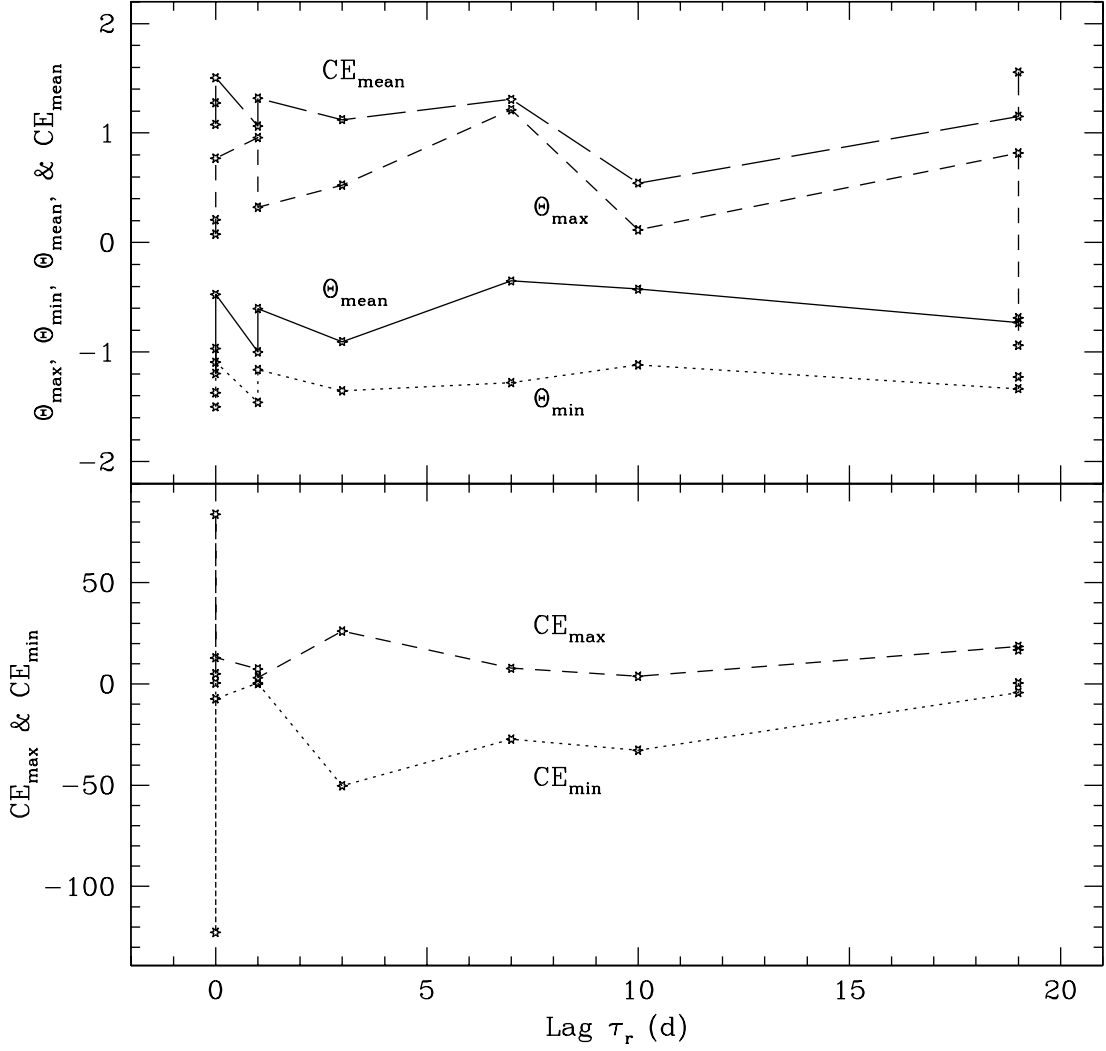}
\caption{Behaviour of $\Theta$ and CE with $\tau_{r}$:  The minimum, maximum, and mean values of both  $\Theta$ and CE during all outbursts are plotted against $\tau_{r}$. All curves are consistent with one another. $\Theta^{lowest}_{min}=-1.50$ is consistent with its ideal lowest value of $-1.57$. Mean CE and mean $\Theta$ vary similarly. It is seen that the maximum magnitude of CE shows instability as $\tau_{r} \rightarrow 0d$, obtained in OB-I type outbursts in LMXBs with small disc. It does not vary too much with $\tau_{r}\gg 0d$, which is attributed to OB-II type outbursts in LMXBs with big disc.}
\end{center}
\end{figure}
\begin{figure}
\begin{center}
\includegraphics[height=18 cm, angle=0]{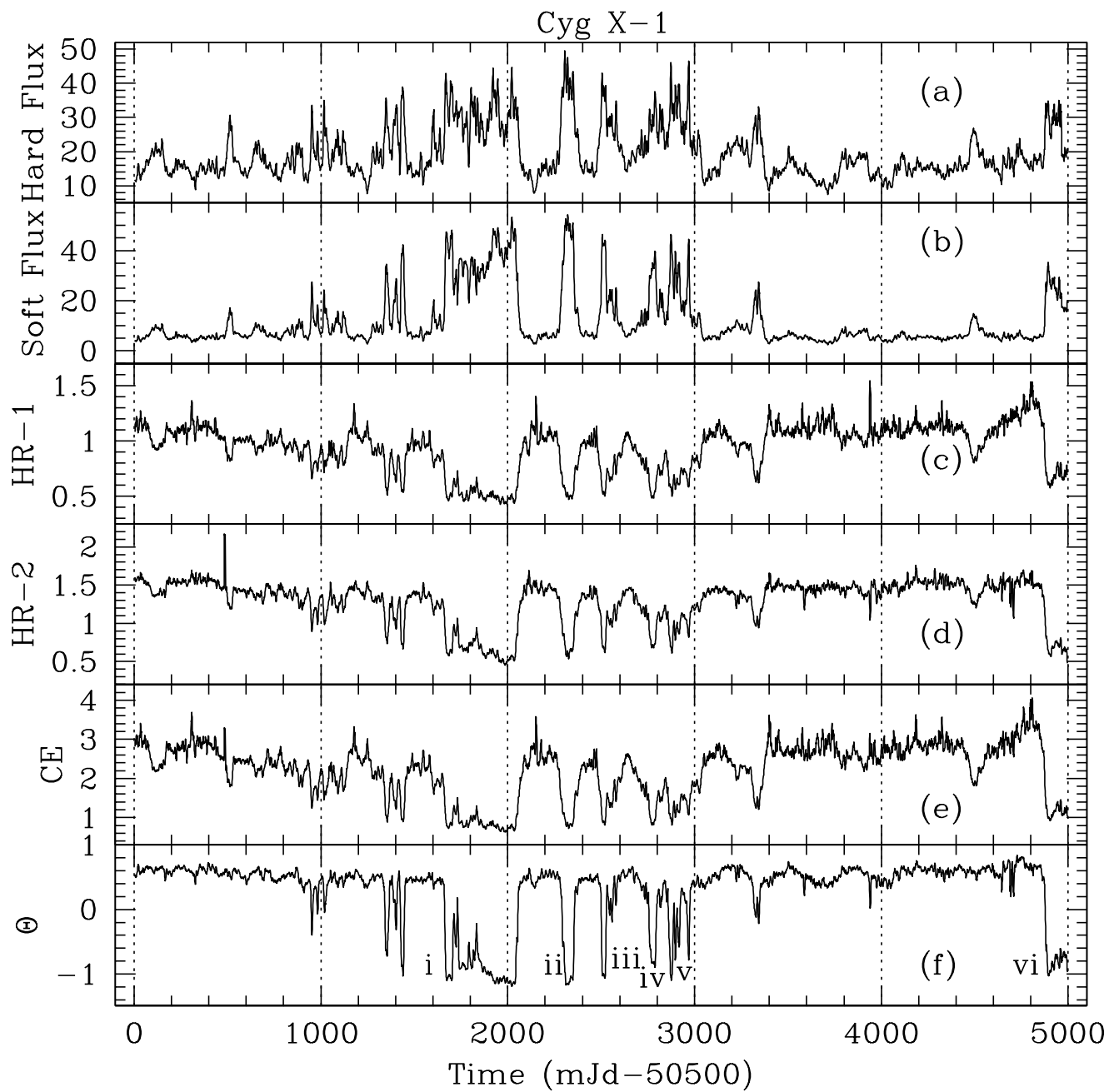}
\caption{Cyg X-1: All curves are plotted with weekly running-averaged long term data for clarity. Hard and soft fluxes are respectively plotted in (a) and (b). (c) to (f) show similar behaviour in HR-1, HR-2, CE, and $\Theta$ respectively. Six chosen flares are marked as i-vi in (f) [see Fig. 11 later].}
\end{center}
\end{figure}
\begin{figure}
\begin{center}
\includegraphics[height=18 cm, angle=0]{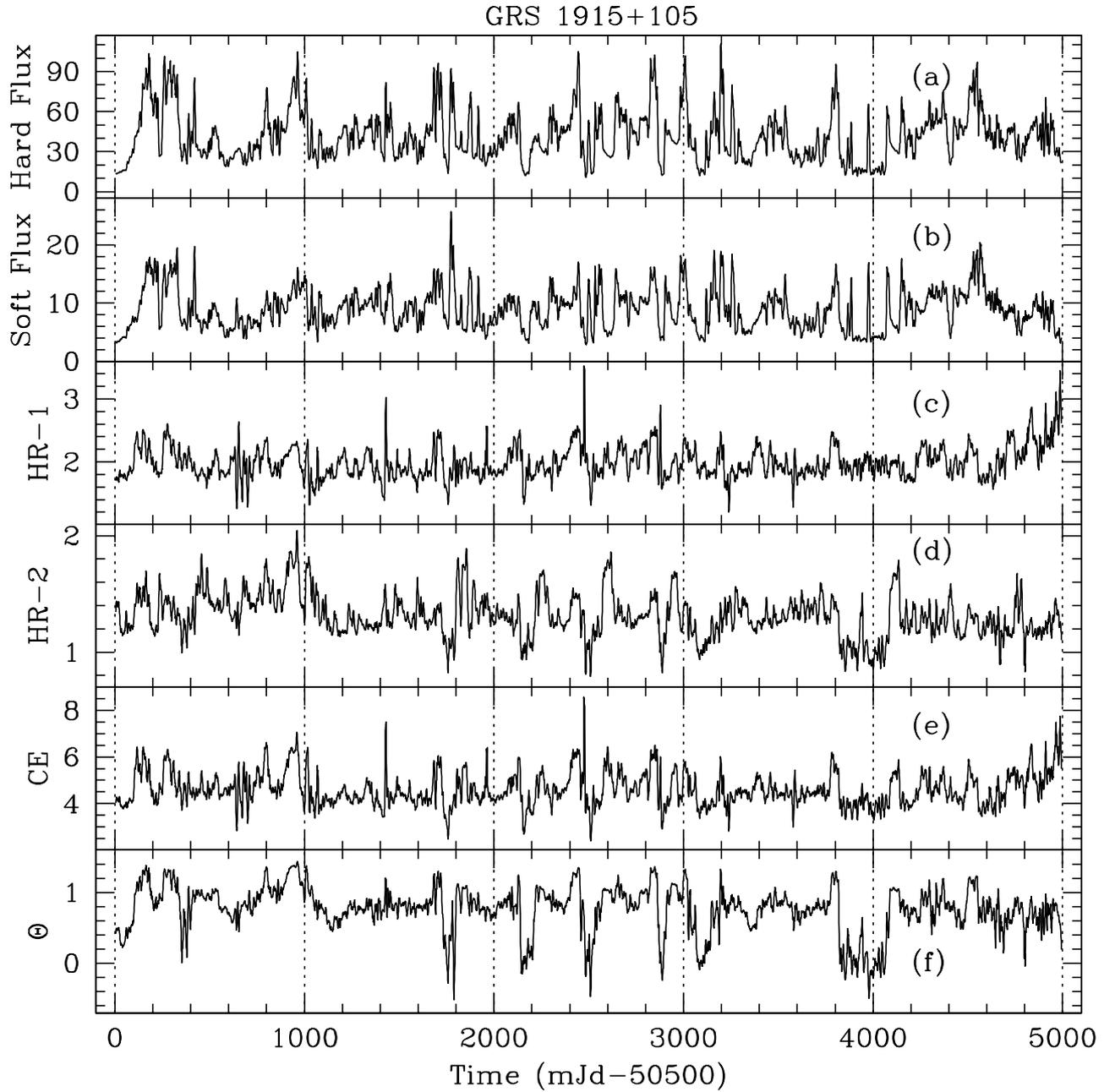}
\caption{GRS 1915+105: All curves are plotted with weekly running-averaged long term data for clarity. Hard and soft fluxes are respectively plotted in (a) and (b). (c) to (f) show similar behaviour in HR-1, HR-2, CE, and $\Theta$ respectively.}
\end{center}
\end{figure}
\begin{figure}
\begin{center}
\includegraphics[height=18 cm, angle=0]{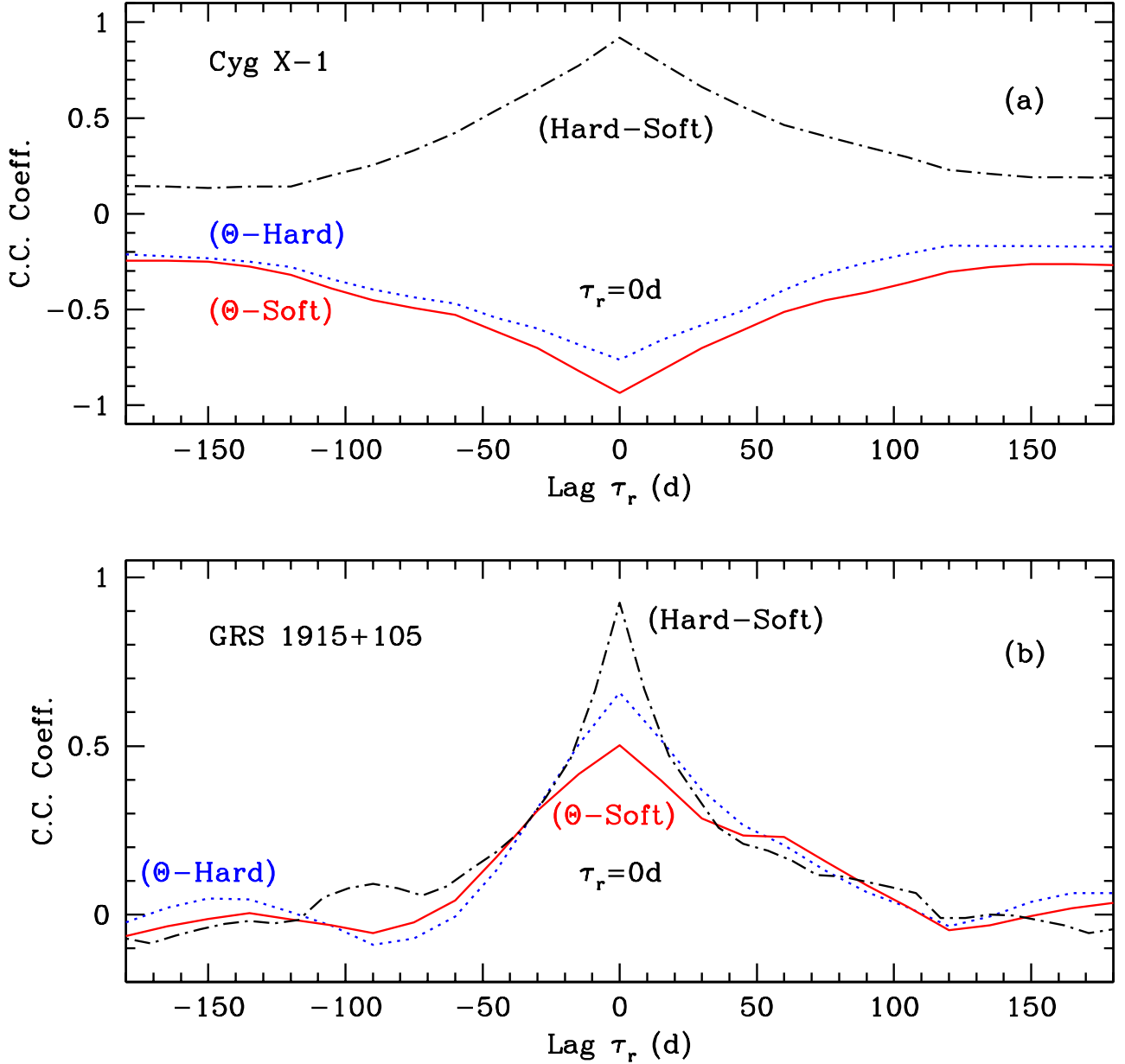}
\caption{Cross-correlation profiles using long-time seconds-data (interpolated at $0.1d$ interval) of Cyg X-1 and GRS 1915+105 are drawn in (a) and (b) respectively. Colour stamps or line types are the same as in Figs. 4-5. No lag is observed in both objects. Behaviour of $\Theta$ with the photon fluxes in Cyg X-1 (a) is similar to those in outbursts in LMXBs.}
\end{center}
\end{figure}
\begin{figure}
\begin{center}
\includegraphics[height=18 cm, angle=0]{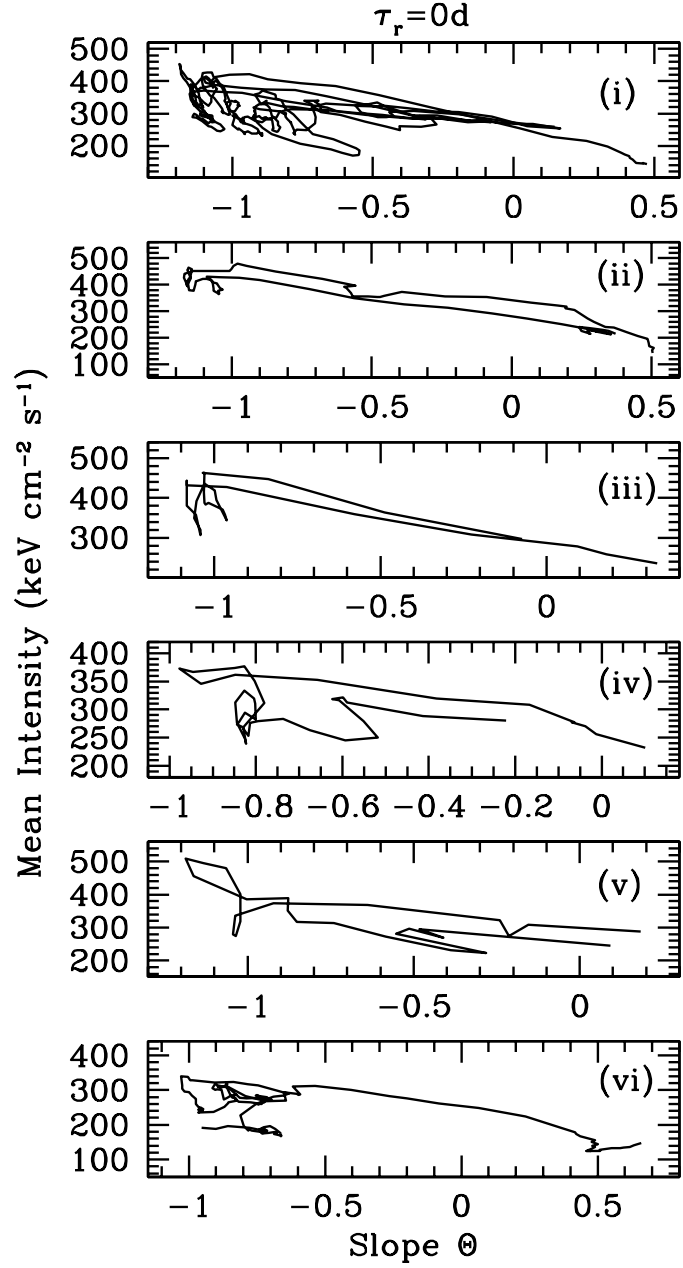}
\caption{Mean intensity in the six flares (marked as i-vi in Fig. 8f) of Cyg X-1 is plotted with $\Theta$. (i) to (v) are similar to the narrow or disrupted {\it hysterisis} loops in outbursts of OB-I type (Figs. 6g-6j); open loop in (vi) signifies partial lack in declining phase (Figs. 8a-b).}
\end{center}
\end{figure}

\end{document}